\documentclass[aip,jcp,twocolumn,reprint]{revtex4-1}

\draft 

\usepackage{CJK}
\usepackage{graphicx}
\usepackage{color}

\usepackage{epstopdf}

\newcommand{\tri}{\triangle}

\newcommand{\sep}{ \ \ \ , \ \ \ }

\newcommand{\beq}{\begin{equation}}
\newcommand{\eeq}{\end{equation}}
\newcommand{\beqn}{\begin{eqnarray}}
\newcommand{\eeqn}{\end{eqnarray}}
\newcommand{\pp}{\partial}
\newcommand{\dd}{{\rm d}}
\newcommand{\ee}{{\rm e}}
\newcommand{\eq}{Eq.\ }

\newcommand{\eqs}{Eqs }
\newcommand{\fig}{Fig.\ }

\newcommand{\cO}{{\cal O}}

\newcommand{\cP}{{\cal P}}

\newcommand{\app}{Appendix\ }
\newcommand{\la}{\langle}

\newcommand{\ra}{\rangle}

\begin{document}


\title{Thermal breakage of a semiflexible polymer: Breakage profile and rate} 



\author{Chiu Fan Lee}
\email[Email:\ ]{c.lee@imperial.ac.uk}
\homepage[homepage:\ ]{http://www.bg.ic.ac.uk/research/c.lee/}
\affiliation{Department of Bioengineering,
Imperial College London, South Kensington Campus,
London SW7 2AZ,
United Kingdom}


\date{\today}

\begin{abstract}
Understanding fluctuation-induced breakages in polymers has important implications for basic and applied sciences. Here I present for the first time an analytical treatment of  the thermal breakage problem of a  semi-flexible polymer model that is asymptotically exact in the low temperature and high friction  limits. Specifically, I provide analytical expressions for the breakage propensity and rate, and discuss the generalities of the results and their relevance to biopolymers.
\end{abstract}


\maketitle 

{\bf Introduction.}
From man-made materials to biopolymers, semi-flexible polymers are ubiquitous in science and engineering. Better understanding of their stability is therefore of high importance. A semi-flexible polymer can naturally be broken by stretching or bending {\it via} external forces, but thermal fluctuations alone will also induce breakage. Fluctuation-induced breakage is particularly relevant in the biological world as many biopolymers are stabilized by hydrophobic interactions and hydrogen bonds between proteins, which are relatively weak compared to covalently bonded synthetic polymers. Indeed, for amyloid fibrils, a kind of polymer implicated in numerous human diseases including Parkinson's and Alzheimer's \cite{dobson_nature03}, it has
 been advocated that thermal breakage could be a key mechanism underlying amyloid fibril proliferation \cite{smith_pnas06,knowles_science09,michaels_jcp14}. Despite the importance of understanding how polymers break, it is surprising that the basic physics remains to be elucidated. For instance, there is currently no concensus on the breakage profile of the polymer \cite{schreck_jpcb13,hong_biophsj13}: some have advocated that breakage happens predominantly in the middle of the polymer \cite{hill_biophysj83} while others have assumed uniform breakage propensity \cite{pallito_biophysj01,smith_pnas06,knowles_science09,szavits_prl14}. This confusion is due in part to the fact that most existing  results rely on considering simplified polymer models in one dimension \cite{sain_pre06,lee_thermal09,ghosh_jcp10} or numerical simulations  \cite{paturej_jcp11}. 
  Here I will provide for the first time analytical expressions of the breakage profile and rate of a highly-rigid polymer in the high friction (highly damped) and low temperature (or high Arrhenius energy) regimes.  


\begin{figure}
\begin{center}
\includegraphics[scale=.49]{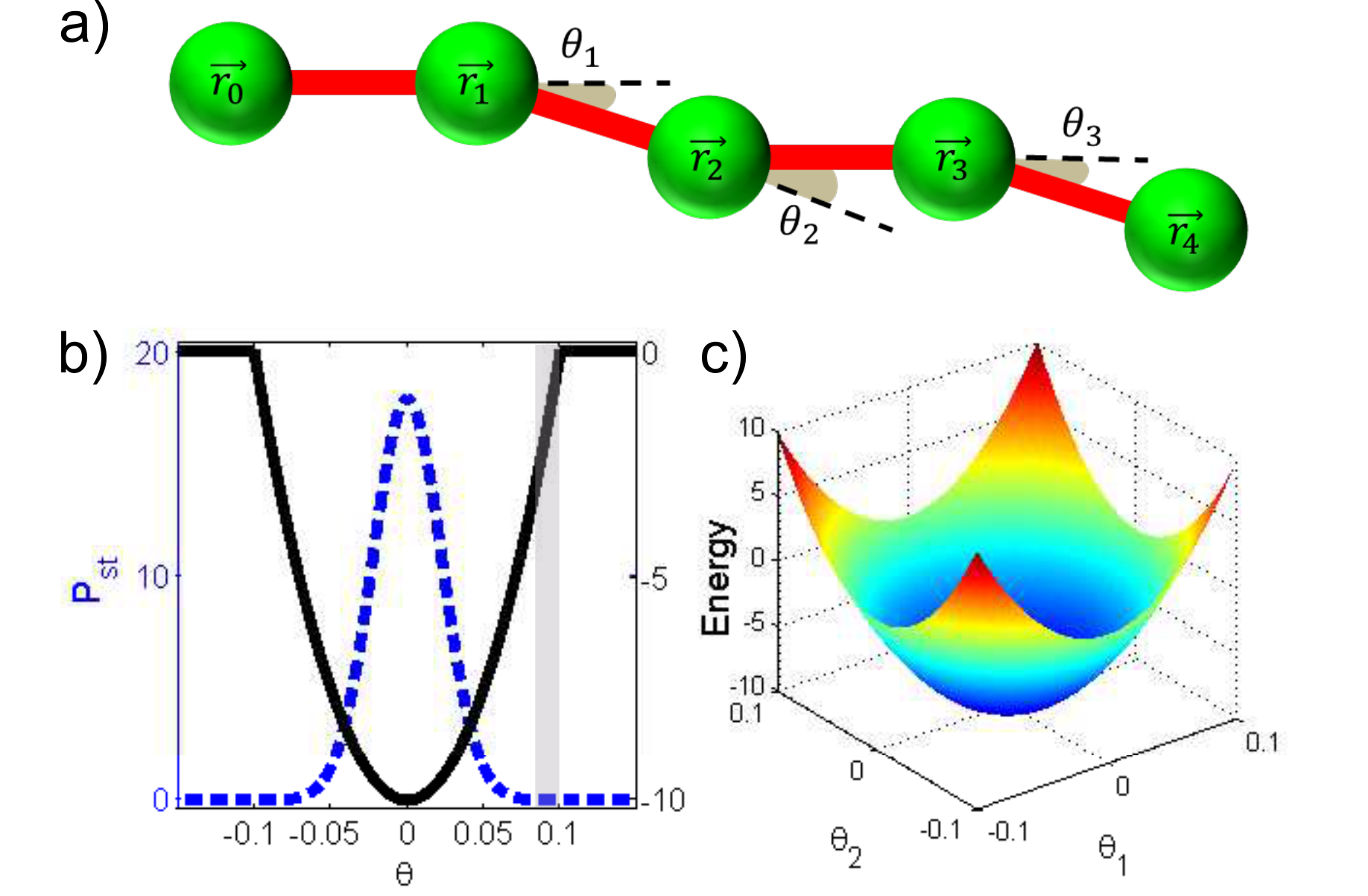}
\end{center}
\caption{a) The bead-and-stick polymer model considered here. b) A particular form of energy function ($U(\theta) = A\theta^2/2$) that enforces bending rigidity (solid blue line) and the resulting quasi-stationary probability distribution $P(\theta) \propto \ee^{-\beta U(\theta)}$, with $\Theta_c=0.1$rad. c)
For a 4-bead polymer, the corresponding energy landscape is two dimensional with 4 minimal-energy breakage points at $(\theta_1,\theta_2) = (\pm \Theta_c,0)$ and $(\pm 0, \Theta_c)$.
}
\label{fig1}
\end{figure}

{\bf Minimal model.}
\label{sec:model}
I will employ a bead-and-stick representation of a semi-flexible polymer  in which the sticks are massless and the beads experience isotropic thermal fluctuations (see \fig \ref{fig1}). 
In the highly damped regime, the equations of motion (EOM) for the beads  are
\beq
\label{eq:eom1}
\frac{\dd \vec{r}_i}{\dd t} = -\frac{1}{\zeta} \vec{\nabla}_{\vec{r}_i}  H + \sqrt{ \frac{2k_BT}{\zeta}} \vec{\eta}_i 
\eeq
where $\zeta$ is the drag coefficient for the beads and $\vec{\eta}_i$ denotes Gaussian noise terms with zero means and unit variance. Also, the tensile and bending rigidities are enforced by two energy potentials $U$ and $V$. Namely,
\beq
H=\sum_{k =1}^{M-1} U(\theta_k) +\sum_{k=1}^M V( r_{k,k-1})
\ ,
\eeq
where $r_{k,k-1} \equiv |\vec{r}_{k,k-1}| \equiv |\vec{r}_{k}-\vec{r}_{k-1}|$ and  $\theta_k \equiv \arccos \left(\frac{\vec{r}_{k,k-1}\cdot \vec{r}_{k+1,k}}{r_{k,k-1} r_{k+1,k}}\right)$.

Thermal fluctuations induce tensile and bending strains on the polymer and the polymer is broken if the bond is stretched or the angle is bent beyond certain thresholds. Since I have previously analysed thermal breakage by fluctuations-induced tensile strain for a polymer on a one dimensional track \cite{lee_thermal09}, I will focus here on breakage by bending, and I will comment on how the results are modified if both breakage by stretching and bending are possible in Discussion. Here, I assume that the polymer is  broken if one of the angles is greater than some material-dependent threshold angle $\Theta_c$ (\fig \ref{fig1}(b)), which is expected to be much less than 1 for highly rigid polymers. With this definition, we are ready to pose the two questions that will be answered in this paper:
\begin{enumerate}
\item
What is the rate for one of the angles to be  pushed beyond $\Theta_c$ due to fluctuations?
\item
What is the breakage propensity as a function of the monomer position in the polymer? 
\end{enumerate}

Since the breakage criterion concerns the set of angles $\theta_k$, it is more natural to consider the EOM of the $\theta_k$ instead of $\vec{r}_k$. Using the chain rule,
\beqn
\frac{\dd \theta_i}{\dd t} &=& \sum_{k=0}^M \vec{\nabla}_{\vec{r}_k} \theta_i \cdot \frac{\dd \vec{r}_k}{\dd t}
\\
\label{eq:theta1}
&=& \sum_{k=i-1}^{i+1} \vec{\nabla}_{\vec{r}_k} \theta_i \cdot  \left(-\frac{1}{\zeta} \vec{\nabla}_{\vec{r}_k}  H + \sqrt{ \frac{2k_BT}{\zeta}} \vec{\eta}_k \right)
\eeqn
where the second equality comes from \eq (\ref{eq:eom1}) and the summation index is now restricted to $[i-1, i+1]$ since variations in other beads do not affect $\theta_i$. We now focus on the highly rigid polymer so that $\Theta_c$ is expected to be small. In this limit, the EOM can be rewritten as  (see \app \ref{app1}):
\beq
\label{eom}
\frac{\dd \theta_k}{\dd t}=\sum_{h=1}^M L_{kh}\left(-\frac{U'(\theta_h)}{\ell \zeta} + \sqrt{\frac{2k_BT}{\ell^2 \zeta}}\xi_h \right)
\eeq
where $\ell$ is the energetically optimal distance between neighboring beads, $L$ is the mobility matrix given in \eq (\ref{eq:L}) \cite{doi_b86}, and the noise terms are now defined by 
 $\la \xi_k(t)  \ra =0$ and  $\la \xi_k(t) \xi_h (t') \ra =2 (L^{-1})_{kh}  \delta(t-t')$.  In other words, the set of angles may be viewed as under the influence of the potential energy $U_{tot}(\{\theta \}) \equiv \sum_{k} U(\theta_k)$ and a thermal heat bath with the nondiagonal mobility matrix $L$.

\begin{figure}
\begin{center}
\includegraphics[scale=.5]{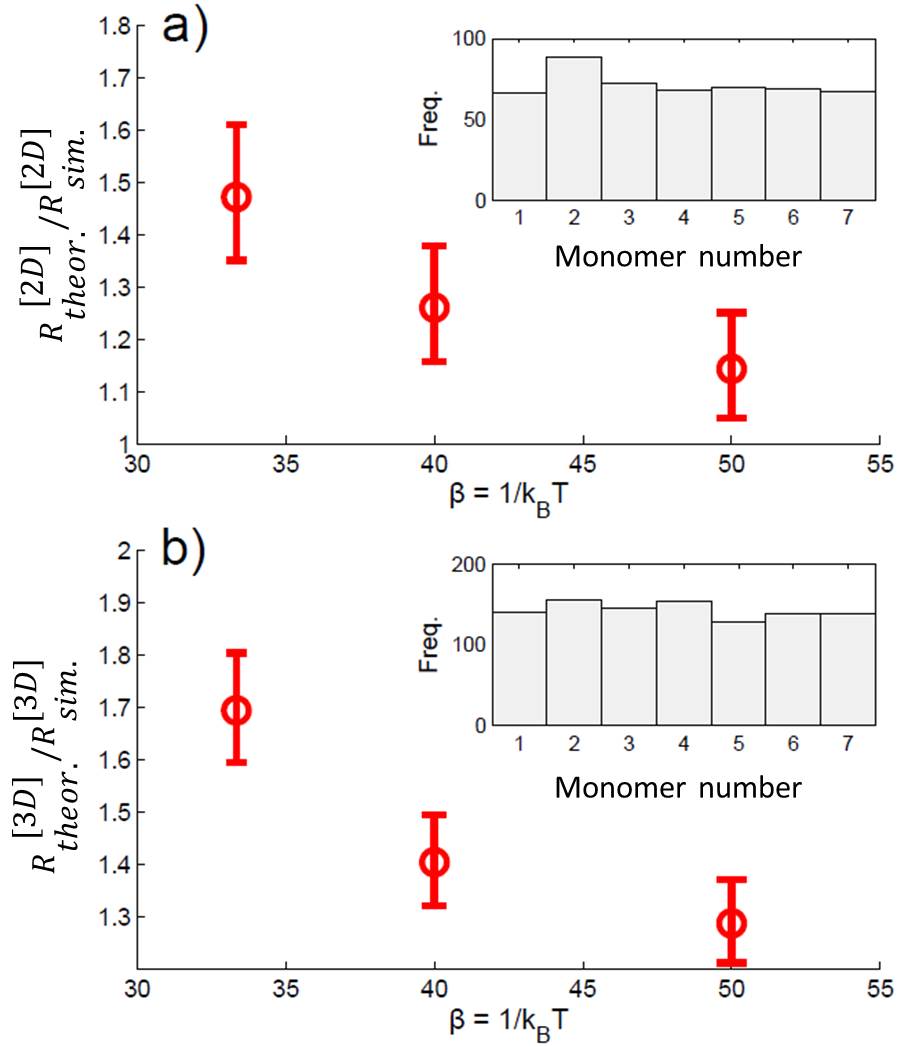}
\end{center}
\caption{Convergence of analytical and simulational results as $T \rightarrow 0$. a) The ratio of the theoretical breakage rate (\eqs (\ref{eq:2D}) \& (\ref{eq:3D}) ) vs.~the breakage rate from Brownian dynamics simulations for a 9-bead polymer in 2D (a) and in 3D (b). The inset plots show the respective breakage frequency with respect to the breakage locations at the lowest temperature considered.
 The energy functions employed and simulation procedure are detailed in the \app \ref{app2}.}
\label{fig2}
\end{figure}

{\bf Two dimensions.}
 \label{sec:2d}
Equipped with the EOM for $\theta_k$, let us now calculate the escape rate and the breakage profile by employing the quasi-static approximation. While the arguments below will be heuristic, the results can be shown to be exact in the asymptotic limit of $\beta \rightarrow \infty$ with $\beta \equiv 1/k_BT$ \cite{matkowsky_siam77,schuss_b09}. In this approximation scheme, we assume that the probability distribution is normalized to one and the distribution is equilibrated to be at the Gibbs state: $P(\{\theta \}) = N_M\ee^{-\beta U_{tot}(\{\theta \})}$ (\fig \ref{fig1}(b) \& (c)). Since at low $T$, the distribution is highly centered around $\theta_k =0$, $P(\{\theta\})$ is well approximated as $N_M\ee^{-\beta A\sum_k\theta_k^2/2}$ where $A=U''(0)$. By integrating this multi-dimensional Gaussian distribution, we find that for a $(M+2)$-bead polymer, $N_M=(\beta A/2\pi   )^{M/2}$. As $T \rightarrow 0$, breakage is highly improbable and is dictated by the configurations where one of the angle is at $\Theta_c$ (say $\theta_1 \sim \Theta_c$) while the rest remain close to $0$.
Around this breakage boundary, we can expand the energy landscape in a Taylor series. As a result, 
the probability distribution $\cP_1\equiv P(\{\theta_1 \sim \Theta_c: \theta_{k>1} \sim 0\})$ can be written as 
\beq
\label{eq:P1}
 N_M\ee^{-\beta \left[A\sum_{k>1} \theta_k^2/2-\tri E + b(\theta_1-\Theta_c)+ \cO((\theta_1-\Theta_c)^2)\right] }
\eeq
where $\tri E= U(\Theta_c) - U(0)$ is the ``Arrhenius'' activation energy term, and $b \equiv  |U'(\Theta_c)|$ is proportional to the gradient of the potential energy at the breakage point (\fig \ref{fig1}). The rate of breakage $R_1$ corresponds to the flux across the breakage boundary $\theta_1 = \Theta_c$. In one-dimension, this amounts to the negation of the diffusion coefficient multiplied by the derivative of the probability distribution. In this multi-dimensional case, the expression is
\beqn
\label{eq:R12D}
R_1&=&- \left(\frac{L_{11}}{\beta \ell^2 \zeta} \right)\left. \frac{\pp }{\pp \theta_1} \right|_{\theta_1=\Theta_c} \int \dd \theta_2 \cdots  \dd \theta_M P_1
\\
&=&
\left(\frac{6}{\beta \ell^2 \zeta} \right)\sqrt{ \frac{ \beta A}{2\pi}} \beta b \ee^{-\beta \tri E} = \frac{3b}{\ell^2 \zeta}\sqrt{\frac{2 \beta A}{\pi }}\ee^{-\beta \tri E} \ .
\eeqn
Since the diagonal elements in the mobility matrix $L$ are identical, we can conclude that \emph{the breakage propensity is uniform along the polymer}. As for the average breakage rate, since there are $M$ angles and there are two breakage position in 2D ($\pm \Theta_c$), the average breakage rate of the polymer is
\beq
\label{eq:2D}
R^{[{\rm 2D}]} = 2MR_1= \frac{6Mb}{\ell^2 \zeta}\sqrt{\frac{2 \beta A}{\pi  }}\ee^{-\beta \tri E}
\ ,
 \eeq
where once again $A =U''(0)$ and $b=|U'(\Theta_c)|$. 
These analytical predictions are supported by Brownian dynamics simulations that show the convergence of theoretical and simulation results as $T \rightarrow 0$.
 (\fig \ref{fig2}(a)).

{\bf Three dimensions.}
 \label{sec:3d}
Going from 2D to 3D introduces a new degree of freedom as the polymer can now rotate around its longitudinal axis. Using the longitudinal axis of the polymer as the $z$-axis, $\theta_k$ corresponds to the polar angle while we denote the azimuth angle by $\phi_k$. Due to this extra degree of freedom, for a $(M+2)$-bead polymer, the normalisation factor at the quasi-static distribution is now modified to 
\beq
N_M =  \left(\int_0^{2\pi}\dd \phi \int_0^\pi \dd \theta \theta \ee^{-\beta A \theta^2/2} \right)^{-M} \simeq \left(\frac{\beta A}{2 \pi}\right)^M
 \ .
\eeq
Once again, defining $\cP_1$ as $P(\{\theta_1 \sim \Theta_c: \theta_{k>1} \sim 0\})$ which has the same expression as in \eq (\ref{eq:P1}), the breakage rate $R_1$ can be similarly calculated as
\beqn
\nonumber
\label{eq:R13D}
R_1&=&- \left(\frac{L_{11}}{\beta \ell^2 \zeta} \right)\left. \frac{\pp }{\pp \theta_1} \right|_{\theta_1=\Theta_c} \int \dd \phi_1  \Theta_c \prod_{k=2}^M (\dd \phi_k \dd \theta_k \theta_k)  \cP_1
\\
&=&
\frac{6 \beta A b \Theta_c}{\ell^2  \zeta}  \ee^{-\beta \tri E}  \ .
\eeqn
Therefore, for a $(M+2)$-bead polymer in 3D, the breakage propensity is again uniform and the overall escape rate is:
\beq
\label{eq:3D}
R^{[{\rm 3D}]}=\frac{6M \beta A b \Theta_c}{\ell^2 \zeta}  \ee^{-\beta \tri E} \ .
\eeq
These analytical predictions are also supported by Brownian dynamics simulation (\fig \ref{fig2}(b)).

{\bf Discussion.}
 \label{sec:dis}
We have seen that for the minimal semi-flexible polymer model considered here, the answers to the two questions posed earlier are analytical tractable in the $T\rightarrow 0$ limit. I will now discuss the generalities and limitations of the  results.

{\it Beyond thermal systems.}
Although the model formulation focuses on polymers at thermal equilibrium, the results remain valid as long as the dynamics of the polymer is well approximated by \eq (\ref{eq:eom1}). In other words, the results also apply to polymers enclosed in a volume that is at ``local'' equilibrium, or under active fluctuations of the form depicted in \eq (\ref{eq:eom1}) \cite{bursac_natmatt05,stuhrmann_pre12}.

{\it Low $T$ limit vs.\ high energy barrier limit.}
Mathematically, these two limits are not reversible: In the first limit, there is only one small parameter in the equation of motions (\eq (\ref{eq:eom1})),  while in the high energy barrier limit, both the fluctuation strength  and the distance from the bottom of the well to the escape boundary become effectively small. Physically, since the units of time and length can be set arbitrarily, the high energy barrier and the low temperature limits are in principle interchangeable with the caveat that the high friction assumption has to remain valid.
This leads to the next comment.

{\it Drag coefficient.} The results presented apply to the high drag regime where the inertia term is neglected. The drag coefficient may be seen as an effective measure of the friction due to {\it both} the internal degrees of freedom within the beads (e.g., protein), and bead-solvent interactions. These two effects combined seem to lead to the general validity of considering biomolecular kinetics in the highly damped regime
\cite{hagen_jpcm05}.

{\it Threshold on bending.} The analysis presented here assumes that the threshold bending angle $\Theta_c$ is small. For biopolymers, I am only aware of one paper on microtubules that provides access to this parameter 
\cite{gupton_currbiol02}. Specifically, it was observed that upon bending, a microtubule forms an arc with curvature of around $1\mu$m$^{-1}$ before breaking 
\cite{gupton_currbiol02}. Since the size of the tubulin  dimers making up the microtubule is around $10$nm, one may deduce that for microtubules, $\Theta_c$ is in the order of $0.01$rad.


{\it Extensile vs.~bending breakage.}
I have omitted  discussion of breakage by extensile fluctuations here because the results have already been derived previously \cite{lee_thermal09}, albeit with the extra assumption that the polymer is on a one-dimensional track. But since the bending fluctuations are orthogonal to the stretching fluctuations. In the low $T$ limit, breakage by stretching is analytically identical for a polymer in 1D or in higher dimensions.
If both extensile and bending breakage events are possible, then the one with a lower Arrhenius activation energy will dominate. If both have the same activation energy, then the relative proportion of bending vs.~tensile breakage will be given by the ratio of the corresponding prefactors  \cite{matkowsky_siam77,schuss_b09}.

{\it Other types of potential energy.} Although our analysis assumes that $U'(\theta_c)$ is non-vanishing here for simplicity. Similar  analysis can be extended to other types of energy such as the single-hump energy function usually employed in the discussion of Kramers escape rate \cite{hanggi_rmp90}. The expression for the breakage rate will naturally be modified accordingly, but the conclusion that the breakage propensity is uniform remains valid.


{\it Internal structure and end effects of a polymer.} 
The model polymer considered  (\fig \ref{fig1}(a)) is certainly a drastically simplified version of a real polymer, but in the spirit of a ghost chain in polymer physics, each bead may be seen as a unit of polymer structure with internal dynamics that are decoupled from the bending dynamics considered here \cite{doi_b86}. If such decoupling is valid, the prediction of uniform breakage propensity should hold true even along the body of the polymer with internal structures. However, the ends of the polymers may be subject to different kinds of energy function. For instance, in the case of protein amyloid fibrils where each fibril is a bundle of filaments, the filaments at the ends may not be as tightly bound and so the bending rigidity at the ends may differ from that in the middle of the fibril. As a result, the breakage propensity close to the ends of the polymer may differ from that in the main body of the polymer.
Another complication is the possibility of having a rugged energy landscape that connects two neighboring beads \cite{lee_pre_elongation09,straub_annrev11}. In this case, the drag coefficient may need to be modified using the theory of diffusion on rugged landscape \cite{zwanzig_pnas88}.

{\bf Summary \& outlook.}
I have considered thermal breakage of a semi-flexible polymer and have obtained analytical results in the highly rigid, highly damped and  low temperature limits.
My calculations indicate that a semi-flexible polymer satisfying these conditions has uniform breakage propensity and analytical expressions of the breakage rates in 2D and 3D were provided. All analytical results were verified by Brownian dynamics simulations. The generalities of the results and their relevance to biopolymers were discussed. 
Future work on this problem will include a thorough analysis of the thermalization kinetics of polymerization since both the equilibrium configuration \cite{cates_jpcm90,lee_jpcm12} and the breakage kinetics are now known. Other interesting directions are the incorporation of higher order corrections into the calculations, and the investigation of  polymer breakage under nonequilibrium and anisotropic fluctuations, such as under shear flows \cite{schneider_pnas07,zhang_science09} and sonications \cite{collins_plosbiol04,trigg_biochemj13}, which constitute two standard experimental procedures in investigating biopolymer self-assembly.

\appendix
\section{EOM for the $\theta_k$}
\label{app1}
Let us first focus on the following terms in \eq (\ref{eq:theta1}):
\beqn
&&\sum_{k=i-1}^{i+1}\vec{\nabla}_{\vec{r}_k} \theta_i \cdot \vec{\nabla}_{\vec{r}_k}  H 
\\
\label{eq:A1}
&=&\sum_{k=i-1}^{i+1}\vec{\nabla}_{\vec{r}_k} \theta_i \cdot \vec{\nabla}_{\vec{r}_k}  \left(\sum_{h =1}^{M-1} U(\theta_h) +\sum_{h=1}^M V( r_{h,h-1})\right)
\\
\nonumber
&=&\sum_{k=i-1}^{i+1} \vec{\nabla}_{\vec{r}_k} \theta_i \cdot \vec{\nabla}_{\vec{r}_k}  \left(\sum_{h =k-1}^{k+1} U(\theta_h) +\sum_{h=k}^{k+1} V( r_{h,h-1})\right)
\ .
\eeqn
The terms of the form $ \vec{\nabla}_{\vec{r}_k} \theta_i \cdot  \vec{\nabla}_{\vec{r}_k} V( r_{h,h-1})$ are of order $\cO(\Theta_c)$,  To see this, consider the particular term $ \vec{\nabla}_{\vec{r}_k} \theta_k \cdot  \vec{\nabla}_{\vec{r}_k} V( r_{k,k-1})$ in 2D. Without loss of generality, assume $\vec{r}_{k,k-1}$ is along the $x$-axis, $\vec{\nabla}_{\vec{r}_k} \theta_k = \ell^{-1}\theta_k \hat{\bf x} + \ell^{-1}\hat{\bf y} +\cO(\theta_k^2)$ and $\vec{\nabla}_{\vec{r}_k} V( r_{k,k-1})=\hat{\bf x}$. Since $\theta_k < \Theta_c$ before breakage, $ \vec{\nabla}_{\vec{r}_k} \theta_k \cdot  \vec{\nabla}_{\vec{r}_k} V( r_{k,k-1})=\cO(\Theta_c)$. Other terms of the same form can similarly be shown to be of the same order, which are small for rigid polymer where $\Theta_c\ll 1$. Physically, what it means is that the fluctuations that contribute to bond extension is orthogonal to the fluctuations contributing to  bending in the small angle limit.  The same arguments apply in 3D.

In \eq (\ref{eq:A1}), we are thus left with the following terms:
\beqn
&&\sum_{k=i-1}^{i+1}\sum_{h =k-1}^{k+1}\vec{\nabla}_{\vec{r}_k} \theta_i \cdot \vec{\nabla}_{\vec{r}_k}  U(\theta_h) 
\\
\nonumber
&=& \sum_{k=i-1}^{i+1}\sum_{h =k-1}^{k+1}U'(\theta_h) \vec{\nabla}_{\vec{r}_k} \theta_i \cdot \vec{\nabla}_{\vec{r}_k}  \theta_h \equiv \sum_{h}L_{ih} U'(\theta_h)
\ ,
\eeqn
where $L$ to $\cO(\Theta_c)$ is a $M\times M$ matrix of the form
\beq
\label{eq:L}
L = \left[
\begin{array}{cccccc}
6 & -4 & 1 & 0& 0 & \cdots
\\
-4 & 6 & -4 & 1 & 0& \cdots 
\\
1 & -4 & 6 & -4 & 1  &\cdots
\\
&  & &  \ddots & &
\\
\cdots &   0&0 & 1 & -4 & 6
\end{array}
\right]
\ .
\eeq

The only remaining terms in \eq (\ref{eq:theta1}) that we have to consider are from the coupled noise terms: $\vec{\nabla}_{\vec{r}_k} \theta_i \cdot \vec{\eta}_k \equiv G_{ik}$. To $\cO(\Theta_c)$, $G$ is a $M\times (M+2)$ matrix of the form
\beq
G = \left[
\begin{array}{cccccc}
-1 & 2 & -1 & 0& 0 & \cdots
\\
0 & -1 & 2 & -1 & 0& \cdots 
\\
0 & 0 & -1 & 2 & -1& \cdots 
\\
&  & &  \ddots & &
\\
\cdots &   0&0 & -1 & 2 & -1
\end{array}
\right]
\ .
\eeq
The matrix $G$ is nondiagonal because there are $M$ variables ($\theta_k$) and $M+2$ noise terms ($\eta_k$). One could therefore reduce the number of the noise terms by 2 by introducing a new set of $M$ noise terms $\xi_k$ with statistics depicted below:
\beqn
\la \xi_k(t) \ra &=& 0
\\
\la \xi_k(t) \xi_k(t') \ra &=& 6 \delta(t-t')
\\
\la \xi_k(t) \xi_{k\pm 1}(t') \ra &=& -4 \delta(t-t')
\\
\la \xi_k(t) \xi_{k\pm 2}(t') \ra &=&  \delta(t-t')
\\
\la \xi_k(t) \xi_{h}(t') \ra &=& 0 \ \ {\rm for}\ h \neq k,k\pm 1, k\pm 2
\ .
\eeqn
In other words, the covariance matrix of $\xi_k$ is exactly the matrix $L$, this explains the EOM in \eq (\ref{eom}) \cite{doi_b86}.

\section{Simulation procedure}
\label{app2}
The simulation is based on numerically integrating the set of stochastic differential equations shown in \eq (\ref{eq:eom1})  as follows:
\beq
\vec{r}_i(t+\tri t) =\vec{r}_i(t) -\frac{\tri t}{\zeta} \vec{\nabla}_{\vec{r}_i}  H (\{ r(t)\})+ \sqrt{ \frac{2k_BT \tri t}{\zeta}} \vec{g}_i 
\eeq
where the entries of the  vector $\vec{g}$ are Gaussian distributed random variables with zero mean and unit variance. The time increment $\tri t$ is set to be $2\times 10^{-6}$, the drag coefficient $\zeta$ is one and the energy function $U$ and $V$ in $H$ are of the form:
\beq
U(\theta) = \frac{A}{2} \theta^2 \sep V(r) = \frac{B}{2} (r-1)^2
\eeq
where $A=60$ and $B=400$. The polymer is always initialized in its lowest energy state and the simulation stops if one of the angles $\theta$ is beyond $\Theta_c=0.1$. Five hundred sample runs are performed at each distinct $k_BT$ for the 2D case (\fig \ref{fig2}(a)) and one thousand sample runs are performed at each distinct $k_BT$ for the 3D case (\fig \ref{fig2}(b)). 


\section*{Acknowledgement}
I thank Pablo Sartori (MPI-PKS) for helpful comments.


%

\end{document}